# Street centrality vs. commerce and service locations in cities: a Kernel Density Correlation case study in Bologna, Italy.

Emanuele Strano[1], Alessio Cardillo[2], Valentino Iacoviello[1], Vito Latora[2], Roberto Messora[3], Sergio Porta[1], Salvatore Scellato[4]

[1] Human Space Lab, DIAP, Politecnico di Milano, Via Bonardi 3, 20133 Milano, Italy. [2] Dipartimento di Fisica e Astronomia, Università di Catania, and INFN Sezione di Catania, Via S. Sofia 64, 95123 Catania, Italy. [3] Copernicana S.I. Srl, Via Matteotti 6/P, 23807 Merate (LC), Italy. [4] Scuola Superiore di Catania, Via San Nullo, 5/i, 95123 Catania, Italy.

__ABSTRACT__    *In previous research we defined a methodology for mapping centrality in urban networks. Such methodology, named Multiple Centrality Assessment (MCA), makes it possible to ascertain how each street is structurally central in a city according to several different notions of centrality, as well as different scales of "being central". In this study we investigate the case of Bologna, northern Italy, about how much higher street centrality statistically "determines" a higher presence of activities (shops and services). Our work develops a methodology, based on a kernel density evaluation, that enhances standard tools available in Geographic Information System (GIS) environment in order to support: 1) the study of how centrality and activities are distributed; 2) linear and non-linear statistical correlation analysis between centrality and activities, hereby named Kernel Density Correlation (KDC). Results offer evidence-based foundations that a strong correlation exists between centrality of streets, especially betweenness centrality, and the location of shops and services at the neighbourhood scale. This  issue is at the heart of the current debate in urban planning and design towards the making of more sustainable urban communities for the future. Our results also support the "predictive" capability of the MCA model as a tool for sustainable urban design.*

## 1. Introduction: understanding the grocer's mantra

*«There are three factors that counts for the success of a grocery: location, location and location».* So the grocer's mantra goes, almost an axiom for all those who are in business with anything that relies on some sort of exchange with the customer/public. If we may easily agree on the principle that location matters, but we are not done with just doing well with our business, the next question is: why location matters? What is the property of a good location that makes it that good? Again, the grocer's answer would be loud and clear: centrality.

Everyone knows that a place which is central has some special features to offer in many ways to those who live or work in cities: it is more visible, more accessible from the immediate surroundings as well as from far away, it is more popular in terms of people walking around and potential customers, it has a greater probability to develop as an urban landmark and a social catalyst, or offer first level functions like theatres or office headquarters as well as a larger diversity of opportunities and goods. That's why central locations are more expensive in terms of real estate values and tend to be socially selective: because such special features make them capable of providing a reasonable trade-off to larger investments than in less central spots of the same urban area. From this point of view, centrality is not exactly just a problem for grocers: because the potential of an urban area to sustain community retail and services is a key-factor in achieving a number of relevant urban sustainability goals in the sub-centers of the nodal information city of the future (Newman and Kenworthy, 1999), from social cohesion to self-surveillance, from liveability to local economy vibrancy, from cyclist-pedestrian friendliness to visual landscape diversity, centrality emerges as one of the most powerful determinant in the hands of urban planners and designers in order to understand how an urban area works and eventually where to address policies of renovation and redevelopment. But there is more.

If one looks at where a city centre is located, she or he will mostly – if not always – find that it sprouts from the intersection of two main routes, where some special configuration of the terrain or some particular shape of the river system or the waterfront makes that place compulsory to pass through. That's where cities begin. Then, departing from such central locations, they grow up in time adding here and there buildings and activities, firstly along the main routes, then filling the in-between areas, then adding streets that provide loop routes and points of return, then, as the structure becomes more complex, forming new central streets and places and keeping buildings around them again. It is an evolutionary process that has been driving the formation of our urban fabrics, the heart of human civilization, through most of the seven millenniums of city history until the dawn of modernity and the very beginning of the industrial age. In short, centrality appears to be somehow at the heart of that kind of "marvellous" hidden order that supports the formation of "spontaneous", organic cities (Jacobs, 1961), which again is an issue of crucial importance in the contemporary debate on the search for more bottom-up, "natural" strategies of urban planning beyond the modernistic heritage.

## 2.  Mapping centrality in urban networks: Multiple Centrality Assessment

Despite the evident relevance of centrality in city life, this issue has been rarely raised to the forefront of urban studies as a comprehensive approach to the subject. Geographers and transport planners have used centrality as a means to understand the location of uses and activities at the regional scale on one side and the level of convenience to reach a place from all others in an urban system (Wilson, 2000): in so doing, they have built their understanding on some hidden assumptions, particularly on a notion of centrality that, putting it shortly, is limited to the following: the more a place is close to all others, the more central.

Since the early eighties "Space Syntax", a methodology of spatial analysis based on visibility and integration (Hillier and Hanson, 1984; Hillier, 1996), opened to urban designers a whole range of opportunities to develop a deeper understanding of some structural properties of city spaces, but such opportunities have been seldom understood, often perceived as a quantitative threat to the creativity embedded in the art of city design.

Basically, the ones who have addressed a specific work on centrality in itself are scientists in fields not related with space, particularly structural sociologists and – more recently – physicists in the "new sciences" of complex networks (Boccaletti et al, 2006). Drawing mainly from those streams, we have recently proposed a whole set of procedures and techniques named Multiple Centrality Assessment (MCA) aimed at the spatial analysis of centralities in urban networks constituted by streets as links or "edges", and intersections as "nodes" (Porta et al, 2006a,b, 2007; Cardillo et al, 2006; Crucitti et al, 2006a,b; Scellato et al, 2006; Scheurer and Porta, 2006). MCA's main characteristics are the utilization of a standard "primal" format of street network's representation, the definition of centrality as a multiple concept described by a set of different peer indices, and the anchoring of all measures on a metric computation of spatial distances along the real street footprint (graph's edge).

While we forward the reader to quoted bibliography for a deeper discussion of such characteristics, it is important here to highlight that MCA's final results are maps of street networks based on the attribution of a centrality "weight" to every segment of the system, which means every street space between one couple of intersections.

## 3. Correlating densities: Kernel Density Estimators in urban analysis

After several first pioneering studies in mathematical statistics (Akaike ,1954; Rosenblatt, 1956; Parzen, 1962), Kernel Density Estimators (KDEs) have been applied to obtain a smooth estimate of a univariate or multivariate probability density from an observed sample of events (Bailey and Gatrell, 1995). Used in many fields that deal with space related problems, from geography to epidemiology, criminology, demography, ethology, hydrology and urban analysis, KDEs are basically a technique aimed at creating a smooth map of density values in which the density at each location reflects the concentration of points in the surrounding area. In spatial analysis, KDEs are therefore a form of surface modelling that attributes values to any location between observed data points (Silverman, 1986).

Such probability techniques of analysis have proven effective in revealing spatial patterns that would not emerge otherwise by the sole distribution of point events, therefore resulting supportive of decision making in public policies related to issues anchored in space: in particular, KDEs are often used for comparative studies of the evolution of spatial events over time or for visually correlating the location of spatial events with other background environmental features, like for instance the presence of front doors (background) with the density of burglaries (events, foreground) in a neighbourhood. A weak point in KDE is the subjectivity in the definition of the bandwidth factor *h*, i.e. the radius of the circular region around each point event (or the width of the buffer around linear events) covered by the density surfaces eventually overlaid on the given cell, that deeply affects the smoothness of the interpolation between observed "event" data (Cao et al, 1994), while both theory and practice suggest that the choice among the various kernel functions does not significantly affects the statistical results (Epanechnikov, 1969).

KDEs have been mostly used in spatial analysis for revealing pattern of clustering of events belonging to the same category, which means to the same layer in a GIS environment. For instance, crime occurrences (Anselin et al, 2000) or street intersections (Borruso, 2003) in urban environments have been investigated by deepening their spatial clustering by means of density estimations. On the other hand, the correlation between phenomena belonging to different categories have been mostly investigated at the condition that they were structurally cross-referenced sharing a field in their respective databases: that was the case of Space Syntax studies on the correlation between street integration, a property of the spatial configuration of urban streets, and the most diverse socio-economic and environmental indicators like pedestrian flows, crime events, retail commerce vitality and pollution (Penn and Turner, 2003). However, a unique potential of KDEs when performed in a Geographic Information System (GIS) environment for urban analysis purposes, lays on the opportunity that it gives to overcome the need of cross-referenced fields for the correlation analysis of *n* different phenomena by simply correlating their proximity in space. Most municipalities in the advanced world have in fact developed a massive amount of information ranging from environmental to economic, demographic or sociologic, that do not share anything but the same geo-referential system: they are, in fact, represented in the same space.

Another unique feature of KDEs in the correlation analysis of space-related variables is the more realistic interpretation of the graduality of spatial influence that they provide, due to their smoothing behaviour: in fact, that seems to capture the experiential notion that, in example, the "effect" of a central street space is not just limited to the curtain facades but instead "spreads" to adjacent spaces and streets to a certain extent, with a decrease in intensity which is a function of distance; or, again, that the "prominence" of one crossing is higher than that of any single converging street: coherently, cells in a crossing proximity are assigned a kernel density value which multiplies that of the converging streets because of the overlaying of the correspondent density surfaces. This feature, which turns out to be essential in the analysis of urban streets, implies the direct implementation of KDE on linear rather than just punctual events: in so doing, the property of a street (centrality in the present research) acts like a "weight" in the computation of that property's density in space.

## 4. Kernel Density in Bologna: distributions of street centrality and commerce/service activities

Bologna is a less than half a million inhabitants important urban centre located in the middle of the river Po plain at the convergence of the main (historical) routes that connect Florence and southern Italy with the northern part of the Country. It is a wealthy city, a relevant political laboratory, the main national transportation hub and the location of the "Alma Mater", the most ancient university in the world.

The aim of the present study on the city of Bologna is to shed light on the correlation between the centrality of streets and the presence of retail commerce and community services. Data were provided by the municipality of Bologna in two separate datasets. The first dataset included all ground floor retail commercial and service activities for the whole urban area stored in an ESRI point shapefile format. For each of the $n$ activities, we were given the location in two dimensional geographic space, that will be indicated in the following as $X_i$, $i=1,2,...n$ (each $X_i$ has to be intended here as a vector in a two dimensional space). For the purpose of this study we then split this dataset into two separated layers, one for retail commercial activities alone ($n_{comm}$=7,257 points), and one for commercial and service activities altogether ($n_{comm+serv}$=9,676 points). The second dataset included the "primal" graph of all streets in the same area stored in a ESRI polyline shapefile format: the dataset counted 7,191 street segments (edges) and 5,448 intersections (nodes). The two datasets were not structurally correlated, but they were coherently geo-referenced such that they resulted perfectly overlayable.

The Kernel Density Correlation (KDC) study proceeded by firstly setting a rectangular region $R$ that encompassed the whole extension of both datasets, and secondly dividing the region in 2,771,956 square cells (edge=10mt). Density has then been calculated, for each of the cells, with reference to events falling within a bandwidth of 100, 200 and 300 meters from the centre of the cell.

In the case of activities, which were represented as points defined in geographic space, a smoothly curved surface was fitted over each point. The surface value was highest at the location of the point and diminished with increasing distance from the point reaching 0 at the circumference of the circular bandwidth from the point; differently from streets, activities were non-weighted entities and, as such, the volume under the surface equalled 1 in all cases. The density of activity at each cell was then calculated by adding the values of all the kernel surfaces where they overlaid the cell's center, following:

$$\hat{f}_h(x) = \frac{1}{nh} \sum_{i=1}^{n} K\left(\frac{x - X_i}{h}\right) \tag{1}$$

In this formula, $K$ denotes the kernel function and $h$ its bandwidth, $x$ represents the position of the centre of each cell, $X_i$ is the position of the $i$-th activity, and $n$ is the total number of activities ($n$ will coincide, in turn, with $n_{comm}$ or with $n_{comm+serv}$). The kernel function $K(y)$ is a function defined for the two-dimensional vector $y$, and satisfying the normalization: $\int_{R^2} K(y)\,dy = 1$. Usually, it is a radially symmetric decreasing function of its argument. The most commonly adopted kernel is the exponential function: $K(y) = (2\pi)^{-1/2} \exp(-\frac{1}{2} y^2)$. In our computation we make use, instead, of the following kernel function, described in Silverman (1986, p. 76, equation 4.5):

$$K(y) = (3\pi)^{-1}(1 - y^2)^2 \qquad \text{if} \quad y^2 < 1$$

$$K(y) = 0 \qquad\qquad\qquad \text{otherwise} \qquad (2)$$

This function has the advantage that can be calculated more quickly than the standard kernel. In fact, with this kernel, if the distance of the activity *i-th* from the centre of the considered cell is larger than the bandwidth *h*, then the activity *i-th* does not contribute to the summation.

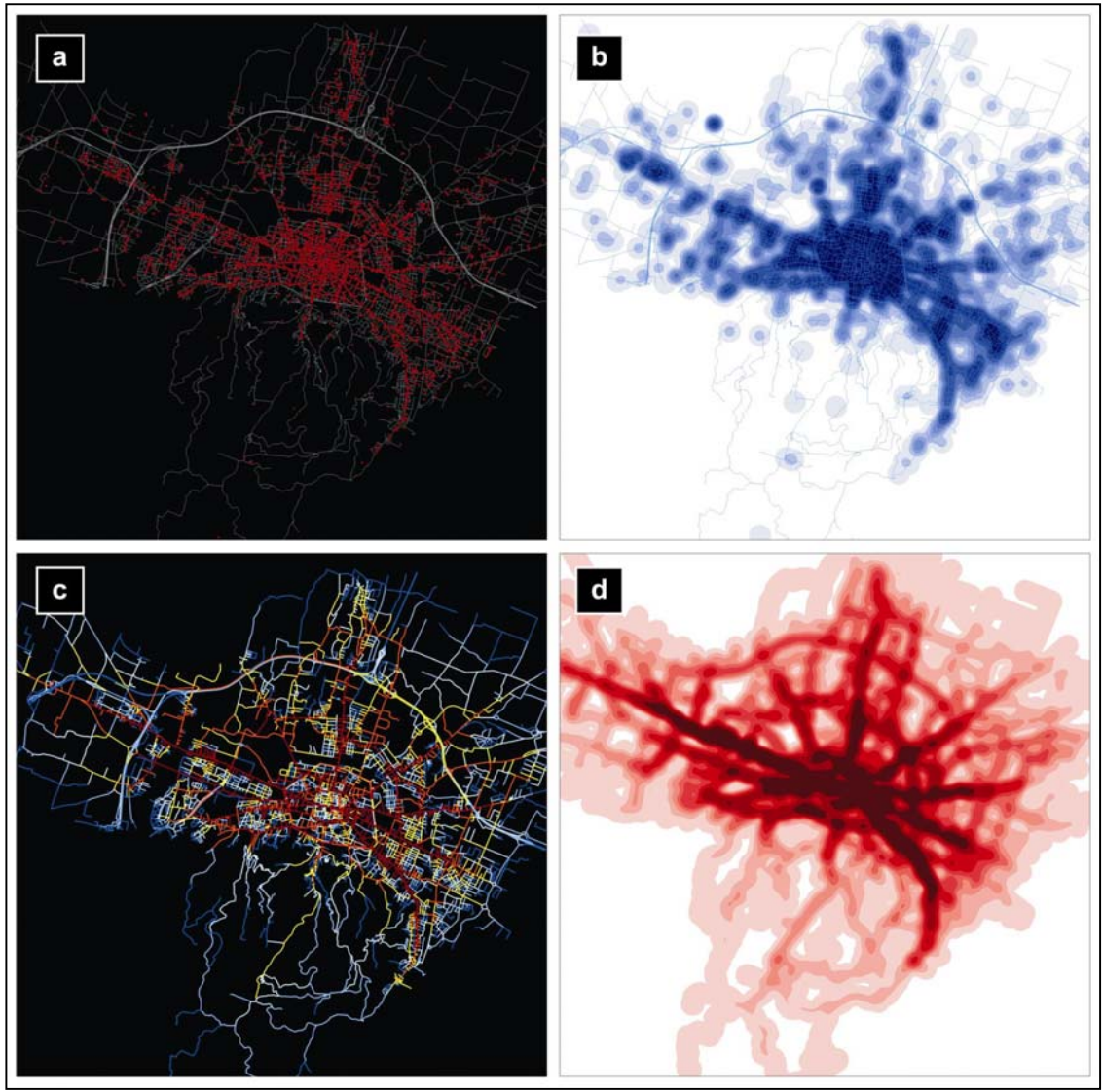

Fig 1.  Kernel Density  in Bologna. Above: (a) the location of ground floor commerce and service activities (red dots) and (b) its kernel density estimation map (bandwidth=300mt). Below: (c) street global betweenness centrality as a result of MCA (colours from blue to red show 9 quantile classes of centrality values from the lowest to the highest) and (d) its kernel density estimation map (h=300mt). An evident visual correlation emerges between the two density maps that is captured statistically by the highest Pearson linear correlation score among the 30 hereby considered (see tab.2).

In the case of streets a smoothly curved surface was analogously fitted over each graph edge. The surface was defined so that the volume under the surface equalled the product of *the portion of line length* included in the bandwidth region and the *centrality value* associated to that edge, on the basis of three different indices of centrality discussed in previous works (Porta et al, 2006a,b; Crucitti et al, 2006a,b). As for the activity case, the density of centrality at each cell was then calculated by adding the values of all the kernel surfaces where they overlay the cell's center. The use of the kernel function for lines is adapted from the same function for point densities quoted above in formula (1) and formula (2).

The output of this first stage of the Kernel Density Correlation process were 21 identical raster layers (tab.1) separately representing the kernel density estimation of the three centrality indices (calculated globally and locally by MCA, excluding betweenness centrality that was calculated only globally), and the two activity categories (commerce and together commerce+services).

| Order # | Centralities | | | KDE Bandwidth |
|---|---|---|---|---|
| | Index | Description | MCA distance factor d | Meters |
| 1 | $C^B_{Glob}$ | Global betweenness centrality | all | 300 |
| 2 | $C^C_{Glob}$ | Global closeness centrality | all | 300 |
| 3 | $C^S_{Glob}$ | Global straightness centrality | all | 300 |
| 4 | $C^B_{Glob}$ | Global betweenness centrality | all | 200 |
| 5 | $C^C_{Glob}$ | Global closeness centrality | all | 200 |
| 6 | $C^S_{Glob}$ | Global straightness centrality | all | 200 |
| 7 | $C^B_{Glob}$ | Global betweenness centrality | all | 100 |
| 8 | $C^C_{Glob}$ | Global closeness centrality | all | 100 |
| 9 | $C^S_{Glob}$ | Global straightness centrality | all | 100 |
| 10 | $C^C_{800}$ | Global closeness centrality | 800 | 300 |
| 11 | $C^S_{800}$ | Global straightness centrality | 800 | 300 |
| 12 | $C^C_{800}$ | Global closeness centrality | 800 | 200 |
| 13 | $C^S_{800}$ | Global straightness centrality | 800 | 200 |
| 14 | $C^C_{800}$ | Global closeness centrality | 800 | 100 |
| 15 | $C^S_{800}$ | Global straightness centrality | 800 | 100 |

| Order # | Activities | | KDE Bandwidth |
|---|---|---|---|
| | Index | Description | Meters |
| 16 | *Comm+Serv* | Retail commerce and community service | 300 |
| 17 | *Comm* | Retail commerce | 300 |
| 18 | *Comm+Serv* | Retail commerce and community service | 200 |
| 19 | *Comm* | Retail commerce | 200 |
| 20 | *Comm+Serv* | Retail commerce and community service | 100 |
| 21 | *Comm* | Retail commerce | 100 |

*Tab 1.  The 27 output layers of Kernel Density Correlation's the first stage applied to the ground floor commerce-services activities and to the street centrality of Bologna urban area.*

The choice of the bandwidth in any KDE application is a well known statistical key-issue (Williamson et al, 1998, Levine, 2004). In this case, all layers have been calculated at three bandwidths ($h$=100, 200 and 300 meters) in order to test both the smoothness performance of KDE at that urban scale and the substantive relevance of such metric thresholds that hold a clear significance in urban studies, being related to concept of street, block and neighbourhood pedestrian sheds. It is worth noting that areas covered by 300 to 400 meters radius circles are usually taken as a reference for a good five minutes walk coverage, and, as such, define the extension of neighbourhood centres, the building blocks of the the "transit-oriented" hierarchical community structure that shapes the sustainable city of the future (Frey, 1999; Calthorpe and Fulton, 2001; Cervero, 2004).

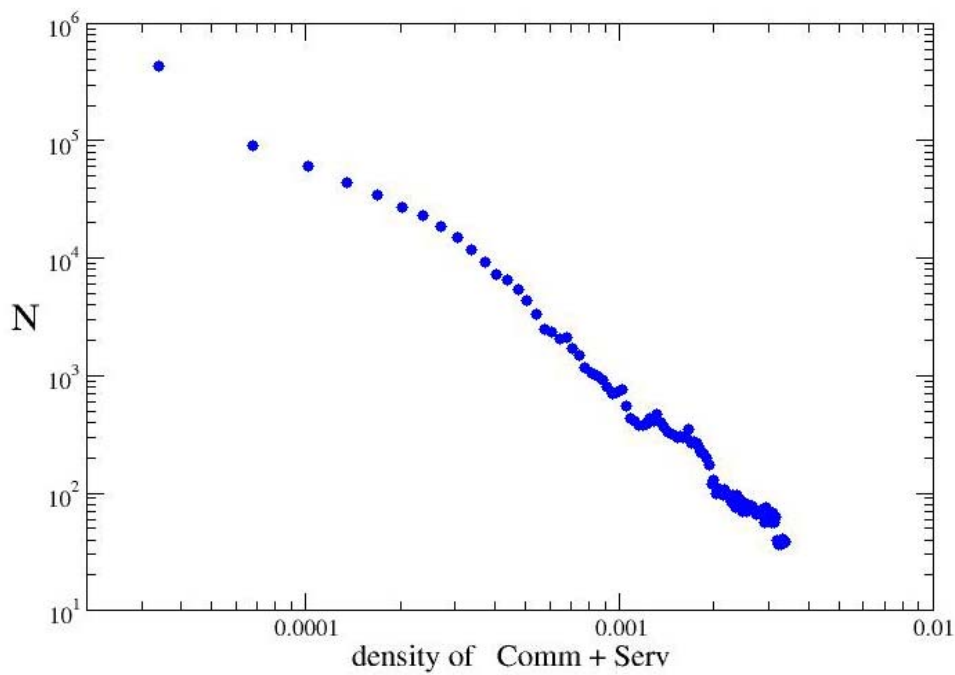

Fig 2.  *Distribution of commerce and service activities in Bologna. We report the histogram of the number of cells with a given density of activities. Densities of activities are evaluated through the Kernel Density method of formula 1, with a bandwidth h=300 meters.*

In this section the focus will be on the statistical distribution of density of both activities and centralities among all the cells in region *R*, while the study of the correlations centrality-activity, the Kernel Density Correlation (KDC), will be discussed in the next section. In order to mathematically quantify the geographic distributions, such as those depicted in panels (b) and (d) of fig.1, we have computed the number of cells with a given density of activities *a*, and the number of cells with a given density of centrality *c*, respectively as a function of *a* and *c*. The results are illustrated in fig.2 and fig.3. In fig.2 we report the number of cells with a density of commerce and service activities in the range *[a,a+Δa]*, as a function of the density value *a*. We have used a bandwidth *h=300m*, and a binning with *Δa=3,396*10⁻⁵*. The log-log scale used in the plot, indicates that the distribution is extremely heterogeneous, and can be rather well approximated by a power-law behaviour (a straight line in the log-log plot) spanning up to two orders of magnitude. This means that, although most of the cells have a small density of activities (of the order of 0.00001-0.0001), there are a few cells with an extremely (even 100 times) larger density of activities.  Or, in other words, the activities are distributed in Bologna in such a way that the average activity density is equal to $0,33332*10^{-4}$, while the standard deviation is equal to $1,4732*10^{-4}$, which is much larger than the average. Therefore, it appears that even the larger bandwidth does not "flatten" the descriptive potential of the process while, quite on the contrary, still captures striking differences in the territorial distribution of activities. Such range of difference is certainly an outcome of the typical concentric shape of the medieval street structure, which favours the concentration of spatial values in a fraction of the regional space; however, we should highlight that the such distribution parallels that exhibited by betweenness centrality (fig.3, panel a). Concentration of spatial centrality as well as accessibility to the functional backbone of community life, therefore, both appear "natural" outcomes of an historical evolution rather than the manifestation of a relatively recent market "distortion".

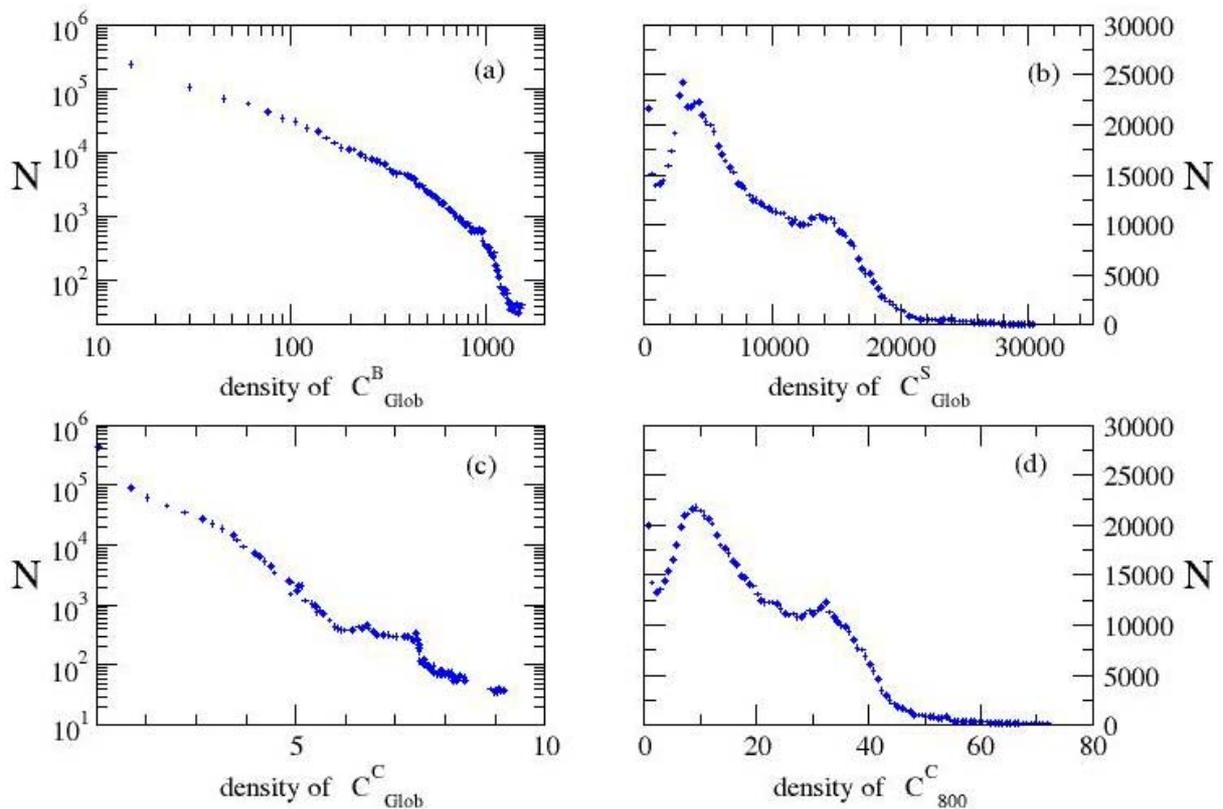

*Fig 3.  Distribution of centralities in Bologna. We report the histogram of the number of cells with a given density of centrality. Densities are evaluated through the Kernel Density method and using a bandwidth h=300 meters. Four measures of centralities from MCA are reported: (a) global betweenness, (b) global straightness, (c) global closeness, (d) local closeness evaluated at a distance of 800 meters.*

In fig.3, in fact, we focus on how centrality indices are distributed. We show only four centrality indices of the 15 calculated by MCA and reported in table 1, namely betweenness (a), straightness (b), closeness evaluated globally (c), and closeness evaluated locally on a scale of 800 meters (d).  In all the four cases, the bandwidth adopted in computing kernel density was of *h*=300 meters, and we plotted the number of cells with a centrality density in a range *[c,c+Δc]* as function of *c*. As for *Δc* we have used the values 45,310 for global betweenness, 476,902 for the straightness and 1,767 for all the closeness. Betweenness is the only one of the centrality measures that exhibits a heterogeneous density distribution, such as that found for the density of activities. In fact, the curve in panel (a)  can be fitted by a power law. The betweenness density in a cell spans values from 10 to larger than 1000,  with over 100,000 cells having a density equal to 10, and only a few hundreds of cells with a density larger than 1000.  The other three centralities, namely $C^S_{Glob}$, $C^C_{Glob}$ and $C^C_{800}$ are more uniformly distributed and characterized by rapidly decreasing tails. This is confirmed by the fact that average and standard deviation are of the same order of magnitude. In particular, the distribution of $C^C_{Glob}$ is very well approximated by a decreasing exponential curve, represented  as a straight line with negative slope in the lin-log plot of panel (c).  Conversely, both $C^S_{Glob}$ and $C^C_{800}$ are characterized by distributions having two peaks at two different values of centrality density. This means that most of the cells preferentially exhibits a centrality value close to one of such two values. A similar behaviour with the presence of two peaks has been found also for straightness and closeness evaluated locally but on different scales.

# 5.  Kernel Density in Bologna: KDC, correlating street centrality and commerce/service activities

The main idea that underpinned our study was that *centrality* acts as a driving force in the formation and constitution of urban structure interacting with the inner laws that link street geography with several *key land uses* like commerce and services activities at the neighbourhood level. The correlation between centrality and such urban activities should therefore be investigated: we then chose to correlated the density of centrality with the density of activities, in so doing overcoming the lack of cross reference information and reaching a more accurate interpretation of the smooth dispersion of relationships in cities by spatial distance. The resulting methodology, that we named Kernel Density Correlation (KDC), turns a well established *combination* of density values on a cell-by-cell basis (Thurstain-Goodwin and Unwin, 2000), in a *correlation* of the same factors.

Therefore, a correlation table between density of centralities and density of activities was produced by means of a dedicated GIS extension. What we did in this phase was to extract from each layer the values of each cell and to build a correlation table where to each cell (record) are attributed the values of that cell in every layer (field) under comparison (fig.4). We then investigated the table in search of linear and non linear statistical correlations coupling centrality (of the kind of *KD_1* in fig.4) and activity (of the kind of *KD_2, KD_3, … KD_n* in fig.4) layers that shared the same bandwidth *h*, which means 30 couples of layers. We have excluded cells that took zero values in both the elements of the couple of layers under scrutiny (the actual universe of cells considered spanned eventually between some 1,500,000 and some 1,800,000 cells). For each couple we have computed the Pearson correlation index. Such index expresses how much two quantities are linearly correlated by giving a number between -1 (most negative correlation) and 1 (most positive); however, the value of the Pearson correlation decreases as the dataset size increases due to statistical fluctuations (Taylor, 1982).

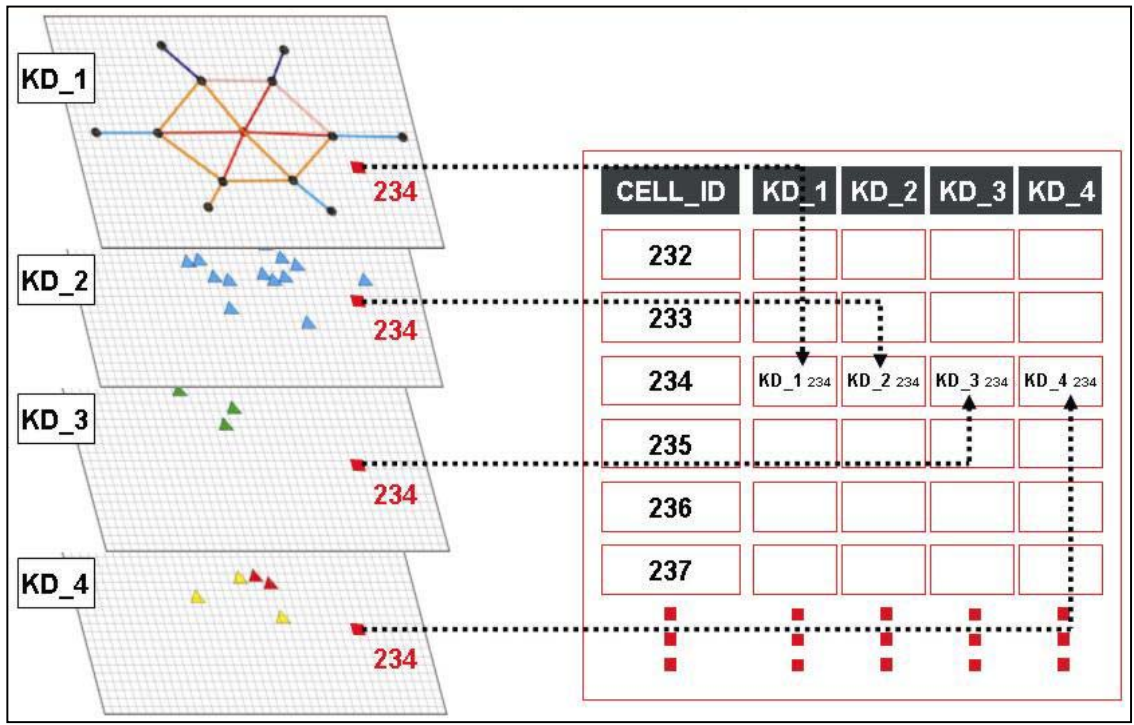

*Fig 4.  A schematic example of the translation of all values that one cell (in this case, cell # 234) takes in a number of KDE layers (produced in stage 1 of Kernel Density Correlation process) into a correlation table. By means of this procedure all kinds of information that are coherently geo-referenced can be correlated statistically with no need of cross-referenced internal structure between different categories (layers) of data.*

The results emerging from the correlation study on the city of Bologna fully confirms the assumption that structural centrality, quantified by the adopted indices, acts as a driving force in the formation and constitution of urban structure, positively influencing the emergence of commerce and services activities at the neighbourhood level. In fact, we found that the first half of the 30 linear correlations hereby investigated between street centrality indices and commerce-service locations (tab.2) takes a Pearson value beyond 0.5, which means, given the large size of correlated datasets, a pretty high positive result. In particular, all the five $C^B_{Glob}$ correlations as well as all the five $C^C_{Glob}$ are included in the first fifteen ranked scores, while $C^S_{Glob}$ is present just twice: that indicates a lower coherence between straightness and activity location when the centrality index is computed at the global level. Global betweenness $C^B_{Glob}$ emerges by far as the highest statistical determinant of commerce and service locations at all scales, with Pearson scores climbing to values well beyond 0.7. This is not unexpected, since betweenness centrality measures the structural centrality of a place by counting the number of times that such place is traversed by the shortest paths connecting couples of places chosen at random on the urban network. In short, the betweenness of a cell is an interpretation of the "traffic" which is present in that cell even if not finding in that cell origin nor destination: it is a property of the space that tells a lot of that "informal" economy based on a widespread presence of secondary functions, i.e. those functions that do not have the power to attract people in themselves, but rather take advantage by the presence of people who are there for other purposes: the core of a community life. Hence a high value of betweenness density in a cell often implies a high value of commerce-service density.

| Rank # | Correlated variables | | KDE Bandwidth | Linear Correlation |
|--------|----------------------|---|---------------|--------------------|
|        | Centralities | Activities | Meters | Pearson index |
| 1  | $C^B_{Glob}$ | Comm+Serv | 300 | 0,727 |
| 2  | $C^B_{Glob}$ | Comm      | 300 | 0,704 |
| 3  | $C^B_{Glob}$ | Comm+Serv | 200 | 0,673 |
| 4  | $C^B_{Glob}$ | Comm      | 200 | 0,653 |
| 5  | $C^C_{Glob}$ | Comm      | 300 | 0,641 |
| 6  | $C^S_{800}$  | Comm+Serv | 300 | 0,620 |
| 7  | $C^S_{Glob}$ | Comm+Serv | 300 | 0,615 |
| 8  | $C^C_{Glob}$ | Comm+Serv | 300 | 0,608 |
| 9  | $C^C_{Glob}$ | Comm+Serv | 200 | 0,583 |
| 10 | $C^B_{Glob}$ | Comm+Serv | 100 | 0,567 |
| 11 | $C^C_{800}$  | Comm+Serv | 300 | 0,565 |
| 12 | $C^B_{Glob}$ | Comm      | 100 | 0,555 |
| 13 | $C^C_{Glob}$ | Comm      | 200 | 0,547 |
| 14 | $C^S_{Glob}$ | Comm+Serv | 200 | 0,546 |
| 15 | $C^C_{Glob}$ | Comm      | 300 | 0,533 |

*Tab 2. Linear correlation (Pearson index) between kernel density of street centrality and kernel density of ground floor activities in Bologna: first 15 positions in ranking.*

In fig.5 we have extended this idea by plotting the density of commerce and services as a function of a cell's centrality. The results are obtained as follows: we divide the cells according to their centrality's density, as we did to build the histograms in fig.3; then, for a given bin of centrality's density, let's say $[c,c+\Delta c]$, we calculate the average density of commerce-activities, where the average is taken over cells with a centrality's density value in the range $[c,c+\Delta c]$, and we plot it as a function of $c$. The results shown in figure, clearly indicate that a higher value of a cell's density of centrality usually implies a higher average density of activites.

The correlation is extremely neat for global betweenness in panel (a) and global closeness in panel (c). Both measures (especially betweenness) are in an almost  linear relationship with the presence of activities. Concerning the global straightness and the local closeness of panels (b) and (d), one observes a steeper increases of the curves. This indicates that, for smaller values of centrality, activities are almost independent from centrality, while the positive influence of centrality over activities is recovered for higher values of centrality. Finally, in the inset of panel (a) we show that correlations are maintained also when betweenness density is evaluated with a bandwidth $h=100m$, although we observe larger fluctuations. This is a further confirmation that the bandwidth value $h=300m$ is to be preferred over $h=100m$ and $h=200m$.

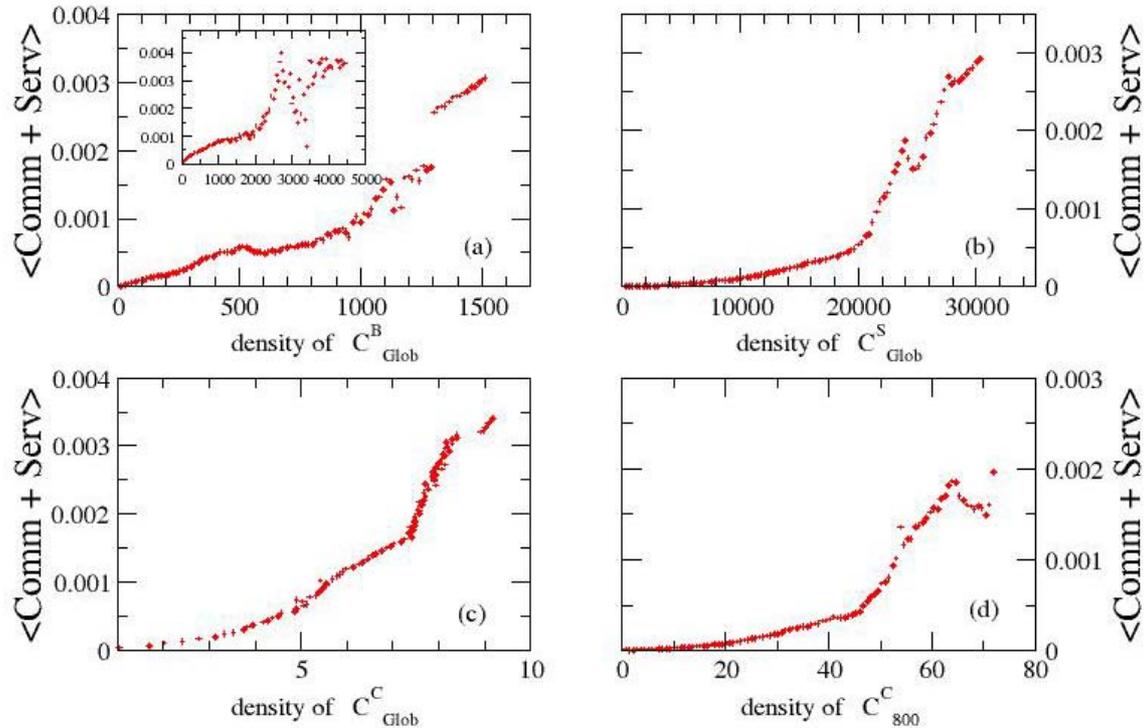

*Fig 5. A graphical plot of the correlation between centrality densities and commerce-service density. The average density of commerce and services is reported as a function of the centrality density: (a) global betweenness, (b) global straightness, (c) global closeness, (d) local closeness are considered. In all the cases the bandwidth for the calculation of the kernel density is set to h=300. Inset in panel (a) shows the correlation between commerce-service density and global betweenness, where now the centrality is evaluated by using a kernel density bandwidth equal to h=100.*

## 6. Conclusions

In this paper the correlation between different measures of street centrality and the presence of ground floor commercial and service activities is investigated in the case study of the city of Bologna, northern Italy. Street centrality has been computed by means of a MCA process, then cross compared with the presence of commerce and service activities by means of a Kernel Density Correlation methodology: in this methodology, centrality and activities have been correlated on the basis of their "inner" density and, in a second step, their mutual proximity in space.

A relevant result of this paper is that activities showed a quite significant orientation to aggregate in the proximity of urban areas where central streets also aggregated, with a particularly high correlation to global betweenness and, to a slightly lesser extent, global closeness centralities. This fully confirms the assumption, within the limits of this case study, that street centrality plays a crucial role in shaping the functional asset of Bologna and that MCA, as a tool for mapping street centrality in cities, captures a most fundamental aspect of the urban phenomenon which plays a substantial role in a wide range of urban planning and design issues.

Another achievement is that the distribution of both densities of betweenness centrality and activities do follow a strong power law behaviour, which means that such territorial resources are "assigned" to space in a quite heterogeneous way: very few places in the city's structure hold a lot of them, while the vast majority hold almost nothing. Such distributions parallel the distribution of centrality in most self-organized complex systems in nature, technology and society, which furthermore confirms our previous studies on the deep "organic" order that seems to have driven the evolution of our historical cities. To what extent this property is due to the concentric structure of the city, an heritage of its prominently medieval street pattern, it is left to further studies.

*Acknowledgements:*
*Streets and activity datasets were kindly provided, on behalf of the Municipality of Bologna, by dr. Andrea Minghetti, dr. Giovanni Fini and dr. Gabriella Santoro. The authors wish to thank Giuseppe Borruso for his many valuable suggestions during the elaboration of the research.*